\documentclass[prb,showpacs,showkeywords,groupaddress,superscriptaddress]{revtex4}
\usepackage{graphicx}

\begin{document}

\title{The road from one hole to the stripe phase}
\author{Fan Yang}
\affiliation{Center for Advanced Study , Tsinghua University,
 Beijing, 100084, China} \email{ fyang@castu.tsinghua.edu.cn}
\author{Su-peng Kou}
\affiliation{Center for Advanced Study , Tsinghua University,
 Beijing, 100084, China} \affiliation{Department of Physics, Beijing Normal
University, Beijing, 100874 ,China}

\begin{abstract}
\begin{center}
{\large Abstract}
\end{center}

In this paper, it is shown how a single stripe and a stripe phase
grow from individual holes in low doping regime. In an effective
low-energy description of the t-J model, {\em i.e.,} the phase
string model, a hole doped into the spin ordered phase will induce
a dipolar distortion in the background [Phys. Rev. B{\bf 67},
115103 (2003)]. We analyze the hole-dipole configurations with
lowest energy under a dipole-dipole interaction and show that
these holes tend to arrange themselves into a regular polygon.
Such a stable polygon configuration will turn into a stripe as the
number hole-dipoles becomes thermodynamically large and eventually
a uniform stripe state can be formed, which constitutes an
energetically competitive phase at low doping. We also briefly
discuss the effect of Zn impurities on individual hole-dipoles and
stripes.
\end{abstract}

\maketitle

\section{Introduction}

The stripe phenomenon is one of many interesting and novel properties
observed in high-$T_c$ cuprate superconductors. Static stripes were first
experimentally found in ${\rm La}_{1.48}{\rm Sr}_{0.12}{\rm Nd}_{0.4}{\rm CuO%
}_4$ \cite{static} by neutron scattering, where narrow elastic magnetic
superlattice peaks located at $(\pi (1\pm 2x),\pi ,0)$ and charge-order
peaks at $(4\pi (1\pm x),0,0)$ are clearly identified at doping
concentration $x=0.118$. This result is interpreted as that the
dopant-induced holes collect in domain walls that separate antiferromagnetic
(AF) antiphase domains. This picture is also supported by the x-ray
diffraction experiments\cite{x-ray}. In ${\rm La}_{2-x}{\rm Sr}_x{\rm CuO}_4$
compounds, people have also tried to use dynamical stripes to explain the
observation by the inelastic neutron scattering \cite{lsco}, where narrow
magnetic peaks at the AF wave vectors, $(\pi (1\pm \varepsilon ),\pi ,0)$
and $(\pi ,\pi (1\pm \varepsilon ),0)$ with $\varepsilon \sim 2x$ at low
energies were found. Somewhat similar incommensurate dynamic magnetic
fluctuations were also reported \cite{ybco} in YBCO compounds. Nuclear
quadruple resonance(NQR)\cite{nqr}, muon spin resonance\cite{usr}, and
magnetic susceptibility measurements\cite{sus} all verify the evidence of
stripe in ${\rm La}_{2-x}{\rm Sr}_x{\rm CuO}_4$. Stripes have also been
observed in the oxygen-doped ${\rm La}_2{\rm CuO}_4$ using nuclear magnetic
resonance (NMR) techniques\cite{nmr}. These experimental results suggest
that the {\em stripe instability} may be extensively present in the cuprates
as a competing order, which contributes to the complexity of the phase
diagram.

The existence of the stripes in a strongly correlated electron system was
actually first predicted \cite{zaanen} by Zaanen and Gunnarsson {\em before}
the experimental discovery. They found the stripe mean-field solution in a
two-band Hubbard model, in which holes doped into the parent antiferromagnet
generally tend to arrange themselves into straight lines aligned parallel to
each other, {\em i.e,} the charged stripes. Since the experimental discovery
of the static stripes in the cuprates, theoretical investigations of the
stripe and stripe-related physics have been conducted very intensively in
the high-$T_{c}$ field. Numerical studies of the t-J model by the DMRG
present conflicting conclusions as to the existence of stripe phases in its
ground state, which might be caused by the strong finite-size effects\cite
{dmrg}. Recent theoretical developments in the stripe physics have been
reviewed in Refs. \cite{zaanen2}.

So far the most of theoretical studies on the origin of the stripes in the
cuprates are either based on phenomenological theories or focused on the
static ones at the mean-field level. To truly understand the microscopic
origin of the stripe phase and its competitive relation with the homogeneous
phases (including superconductivity), one needs to know when a stripe can be
melt and be broken into pieces, namely,{\em \ }what it is made of, and when
it can become stable against various kinds of fluctuational effects.

Recently, it has been shown \cite{kou,kou1} that there exists a {\em more}
stable elementary object, known as a hole dipole, in the low-doping spin
ordered phase described by the t-J model. Such a charge $+e$ entity can be
regarded as a dipole composed of a charged vortex (centered at a spinless
holon) and a neutral antivortex which is self-trapped in real space. Due to
the so-called phase string effect, an infinite (logarithmic divergent)
energy is needed if one is to \textquotedblleft destroy\textquotedblright\
such a composite by separating two poles of the dipole infinitely far away.

On the other hand, since each hole dipole is self-trapped in real space, its
kinetic energy is suppressed. Thus the potential energy (from impurities,
for instance) and the dipole-dipole interaction between two holes will
become dominant. In the absence of disorder or impurities and without
considering the long-range Coulomb repulsion, an inhomogeneous instability
has been found in such a system and in particular various stripe
instabilities were suggested \cite{kou1} to occur. In other words, if a
stripe does form in this system, hole-dipoles described above will become
the elementary building blocks. Consequently the fluctuations and dynamics
of stripes as well as the melting of them may be understood and
mathematically described from a new angle based on the hole-dipoles.

In this paper, we shall follow up the stripe instability pointed out in Ref.
\cite{kou1} and demonstrate mathematically how a {\em stable} stripe can
grow from individual hole-dipoles by starting with only a few of them. We
find that these finite number of hole-dipoles generally form a regular
polygon with a minimized potential energy, which is stable against the
perturbations. With the increase of the hole number, the polygon eventually
evolves into a stripe and then stripes as the hole concentration becomes
finite in the thermodynamic limit, which result in the stripe phase. We
further consider the Zn impurity effects on both hole dipoles and stripes
and predict that stripes can be easily destroyed in the presence of random
zincs.

\section{ The Model}

\subsection{The phase-string model}

We start with the two-dimensional t-J model. At half-filling, it turns into
the Heisenberg model with a good description of the magnetic properties in
the cuprates. The Marshall sign rule is found \cite{marshall} in the ground
state of such a model, where the flips of two antiparallel spins at opposite
sublattice sites are always accompanied by a sign change in the ground-state
wave function: $\uparrow \downarrow \rightarrow (-1)\downarrow \uparrow $.
Upon doping, however, this Marshall sign rule will get frustrated by the
motion of the doped holes. When a hole hops from site to site, a sequence of
$\pm $ signs will be left behind which cannot be repaired by spin-flip
processes and is called phase-string \cite{string1}.

The phase-string theory is developed to accurately handle the phase-string
effect in the t-J model. This theory is based on a new kind of
slave-particle formula in which electron operator reads \cite{string1}
\begin{equation}
c_{i\sigma }=h_{i}^{\dagger }b_{i\sigma }(-\sigma )^{i}e^{i\hat{\Theta}%
_{i\sigma }},
\end{equation}
where $h_{i}^{\dagger }$ is a bosonic \textquotedblleft
holon\textquotedblright\ creation operator and $b_{i\sigma }$ is a bosonic
\textquotedblleft spinon\textquotedblright\ annihilation operator,
satisfying the following no double occupancy constraint
\begin{equation}
h_{i}^{\dagger }h_{i}+\sum_{\sigma }b_{i\sigma }^{\dagger }b_{i\sigma }=1.
\end{equation}
Here the nonlocal phase factor $e^{i\hat{\Theta}_{i\sigma }}$ precisely
keeps the track of the singular part of the phase string effect as well as
the fermionic statistics of the electron operator$,$ as defined by
\begin{equation}
e^{i\hat{\Theta}_{i\sigma }}=e^{\frac{i}{2}\left[ \Phi _{i}^{b}-\sigma \Phi
_{i}^{h}\right] },
\end{equation}
with
\begin{equation}
\Phi _{i}^{b}=\sum_{l\neq i}\mathop{\rm Im}\ln (z_{i}-z_{l})\left(
\sum_{\alpha }\alpha n_{l\alpha }^{b}-1\right) ,~
\end{equation}
and
\begin{equation}
\Phi _{i}^{h}=\sum_{l\neq i}\mathop{\rm Im}\ln (z_{i}-z_{l})n_{l}^{h}~.
\end{equation}

The effective phase-string model of the t-J Hamiltonian is given by
\begin{equation}
H_{eff}=-t_{h}\sum_{\langle ij\rangle }\left[ \left( e^{iA_{ij}^{s}}\right)
h_{i}^{\dagger }h_{j}+H.c.\right] -J_{s}\sum_{\langle ij\rangle \sigma
}\left[ \left( e^{i\sigma A_{ij}^{h}}\right) b_{i\sigma }^{\dagger
}b_{j-\sigma }^{\dagger }+H.c.\right] ,  \label{heff}
\end{equation}
with $t_{h}\sim t$, $J_{s}\sim J$. The most important and unique structure
of the phase-string theory is a mutual dual relation in (\ref{heff}): For
holons, a spinon simply behaves like a $\pm \pi $ flux-tube and for spinons,
a holon also behaves like a $\pi $ flux-tube, which are described by the
lattice gauge fields $A_{ij}^{s}$ and $A_{ij}^{h}$ as follows:
\begin{equation}
\sum_{C}A_{ij}^{s}=\frac{1}{2}\sum\limits_{\sigma ,l\in C}(\sigma n_{l\sigma
}^{b}),
\end{equation}
and
\begin{equation}
\sum_{C}A_{ij}^{h}=\frac{1}{2}\sum_{l\in C}n_{l}^{h},
\end{equation}
for a closed path $C$ with $n_{l\sigma }^{b}$ and $n_{l}^{h}$ denoting
spinon and holon number operators, respectively.

\subsection{Holes as dipoles}

In the phase-string theory, the spin-flip operator is defined as
\begin{equation}
S_i^{+}=(-1)^ib_{i\uparrow }^{\dagger }b_{i\downarrow }\exp \left[ i\Phi
_i^h\right],
\end{equation}
with
\begin{equation}
\Phi _i^h=\sum_{l\neq i}\mathop{\rm Im}\ln (z_i-z_l)n_l^h~.
\end{equation}
In the AF spin ordered phase, the spinons are Bose condensed, {\em i.e.,} $%
<b_{i\sigma }>\neq 0,$ with the spins lying in the xy plane. The
polarization direction of the spin ordering is determined by $\left\langle
S_i^{+}\right\rangle =(-1)^i<b_{i\uparrow }^{\dagger }><b_{i\downarrow
}>\exp \left[ i\Phi _i^h\right] \label{s+^}$. From this, we can see that
besides the sign $(-1)^i$, which reflects the staggered AF order, there is
an additional phase $\exp \left[ i\Phi _i^h\right] $ introduced by holons,
which represents a twist of the spins with respect to each holon. Namely,
each time when one circles around a holon once, a $2\pi $ rotation is found
in the direction of the spin ordering. The resulting spin configuration is
called a meron (spin vortex) [Fig.2 of Ref. \cite{kou1}]. A meron costs an
energy which is logarithmically dependent on the size of the system. And for
two holons, the induced spin twists are in the same way such that there
exists a repulsive interaction between them. In order to remove such an
unphysical energy divergence, an antimeron should be induced \cite{kou1}
near every holon-meron to cancel out the spin twists at large distance. An
antimeron is defined by:
\begin{equation}
b_{i\sigma }\rightarrow \tilde{b}_{i\sigma }\exp \left[ i\frac \sigma 2%
\vartheta _i^k\right],  \label{bmeron}
\end{equation}
where $\vartheta _i^k=\mathop{\rm Im}\ln (z_i-z_k^0).$ Here $z_k^0$ denotes
the coordinate of the center of an antimeron labelled by $k$. As a result

\begin{equation}
\left\langle S_{i}^{+}\right\rangle \rightarrow (-1)^{i}\left\langle \tilde{b%
}_{i\uparrow }^{\dagger }\right\rangle \left\langle \tilde{b}_{i\downarrow
}\right\rangle \exp \left[ i\Phi _{i}^{h}-i\vartheta _{i}^{k}\right].
\label{s+}
\end{equation}
Define
\begin{equation}
\phi _{i}^{k}=\Phi _{i}^{h}-\vartheta _{i}^{k}\text{{}}  \nonumber \\
=\text{Im}\ln \text{ }\frac{z_{i}-z_{k}/2}{z_{i}+z_{k}/2}  \label{phik}
\end{equation}
to describe the spin twist, with $z_{k}\equiv e_{k}^{x}+ie_{k}^{y}.$ Here
the meron and anti-meron are centered at $\pm \frac{{\bf e}_{k}}{2},$
respectively. At $|{\bf r}_{i}|>>|{\bf e}_{k}{\bf |,}$ one obtains a dipolar
twist
\begin{equation}
\phi _{i}^{k}\simeq \frac{\left( {\bf \hat{z}\times e}_{k}\right) {\bf \cdot
r}_{i}}{|{\bf r}_{i}|^{2}}.  \label{dipole}
\end{equation}
The energy cost of such a dipole configuration is given as the following
\cite{kou1}

\begin{eqnarray}
{\cal E}_k^d &\simeq &\frac{J_s\rho _c^sa^2}4\int d^2{\bf r}\frac{\left|
{\bf e}_k\right| ^2}{\left| {\bf r}-{\bf e}_k/2\right| ^2\left| {\bf r}+{\bf %
e}_k/2\right| ^2}\text{ }  \nonumber \\
&\simeq &q^2\ln \frac{\left| {\bf e}_k\right| +a}a,\text{\qquad }q^2=\pi
J_s\rho _c^sa^2,\text{~}  \label{edipole}
\end{eqnarray}
which is finite. In the above, $\rho _c^sa^2=\langle b^{+}\rangle ^2$ ($a$
is the lattice constant). This meron-antimeron spin configuration is called
a hole-dipole \cite{kou,kou1}. The displacement connecting the centers of a
meron and an antimeron can be defined as the dipole moment here. The dipole
moment is determined by the two dimensional Coulomb gas theory by a standard
KT\ renormalization group (RG) method\cite{kou,kou1}. Near half filling $%
x\rightarrow 0$ we estimate the centers of a meron and an antimeron as the
lattice constant $\left\langle |{\bf e}_k|\right\rangle =r_0\sim a$.

At the end of this part, we note that in somewhat different contexts, the
concept of hole dipoles has been also suggested by different authors through
different approaches. It is raised in references\cite{aharony1,aharony2}
that the doped holes introduce a local ferromagnetic exchange coupling
between their neighboring ${\rm Cu}^{2+}$, which brings frustration to the
background antiferromagnetism. The frustrating bond acts like a magnetic
dipole. From this picture, they studied the suppression of antiferromagnetic
correlations by the hole-dipoles, and the magnetic phase diagram was
obtained. It is also pointed out in reference\cite{spin bag} that the doped
holes in the form of hole-spin-polaron interact with each other through
dipolar potential. It is just the excitation of these spin-polarons that
forms the stripe\cite{polaron1,polaron2}. In reference\cite{analogy}, an
analogy was given between the doped holes in the AF background with the $%
{\rm He}^3$ impurity in the liquid ${\rm He}^4$. As a result, it is also
found that the mobile-hole creates a long-range dipolar spin-backflow. And
in reference\cite{coulumn}, the interaction between the holes is more
carefully considered, including the long-rang Coulomb interaction, the
dipolar potential, and the short-range attraction. The competing of these
interactions leads to the complicated phase diagram, which includes the
diagonal stripe.

\section{Optimal configurations for multi-hole-dipoles}

A hole-dipole is self-localized \cite{kou1} in space with the suppression of
its kinetic energy. If there is an impurity, such a hole-dipole can be
easily trapped around, whose effect will be discussed in Sec. IV. In this
section, we shall consider the impurity-free case, in which the hole dipole
can be located anywhere in real space due to the translational symmetry. For
multi-number hole-dipoles, the dipole-dipole interaction will determine the
spatial configuration of them. In the following we start with the case of
two hole-dipoles first.

\subsection{Energy-minimal configuration for two hole-dipoles}

For two dipoles well separated from each other, the interaction energy is
given by \cite{kou1}

\begin{eqnarray}
V_{kk^{^{\prime }}}^{d-d} &\simeq &\frac{2q^2}{\left| {\bf r}_{kk^{^{\prime
}}}\right| ^2}\left[ {\bf e}_k\cdot {\bf e}_{k^{^{\prime }}}-2\frac{({\bf e}%
_k\cdot {\bf r}_{kk^{^{\prime }}})({\bf e}_{k^{^{\prime }}}\cdot {\bf r}%
_{kk^{^{\prime }}})}{\left| {\bf r}_{kk^{^{\prime }}}\right| ^2}\right]
\text{ }  \nonumber  \label{vdd} \\
&=&-\frac{2q^2\left| {\bf e}_k\right| \left| {\bf e}_{k^{\prime }}\right| }{%
\left| {\bf r}_{kk^{\prime }}\right| ^2}\cos \left( \varphi _k+\varphi
_{k^{\prime }}\right)  \nonumber \\
&\simeq &-2q^2\frac{r_0^2}{\left| {\bf r}_{kk^{\prime }}\right| ^2}\cos
\left( \varphi _k+\varphi _{k^{\prime }}\right) ,
\end{eqnarray}
in which the size of the dipoles is fixed by $r_0.$ The alignments
of two dipoles are shown in Fig.1.

\begin{figure}[tbp]
\begin{center}
\includegraphics{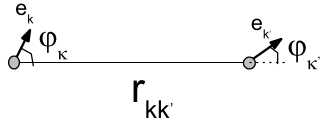}
\end{center}
\caption{The variables in formula(16) is shown. For two dipoles
indexed by k and $k^{^{\prime }}$, $r_{kk^{^{\prime }}}$ denotes
the distance between
their centers, $e_k$ and $e_{k^{\prime }}$ denote their moments, and $%
\varphi _k$ and $\varphi _{k^{\prime }}$ denote the angles between their
moments and the line which connects their centers.}
\label{fig1}
\end{figure}

Under the interaction (\ref{vdd}), two dipoles will adjust their dipole
moment directions to arrive at the lowest energy. It is easy to see that the
condition to minimize their potential energy at a fixed distance is
\begin{equation}
\varphi _k+\varphi _{k^{^{\prime }}}=0,\text{ \ {\rm modulo }}2\pi,
\end{equation}
which results in an attractive potential energy,
\begin{equation}
V_1^{d-d}\sim -\frac{2q^2r_0^2}{\left| {\bf r}_{kk^{\prime }}\right| ^2}.
\end{equation}
Two dipoles will then move closer and closer until reach a least distance $%
2r_0=2\left| {\bf e}_{k^{\prime }}\right| $, determined essentially by the
size of the dipoles. When the distance between two dipoles are near $2r_0 $,
the potential described by (\ref{vdd}) usually is no longer correct.
However, we shall use the formula (\ref{vdd}) approximately at $\left| {\bf r%
}_{kk^{\prime }}\right| \gtrsim 2r_0$ and take the positions of the dipoles
as continuous variables by ignoring the discrete lattice sites in the
following considerations of $n$-dipole case. {\em Then the problem is
reduced to a mathematical one to search for the optimum configuration for
the $n$-dipoles which interact with each other through (\ref{vdd}), under
the constraint that the distance between any two dipoles should be no less
than twice of the average dipole moment $\left| 2{\bf e}_{k^{\prime
}}\right| =2r_0$.}

\begin{figure}[tbp]
\begin{center}
\includegraphics{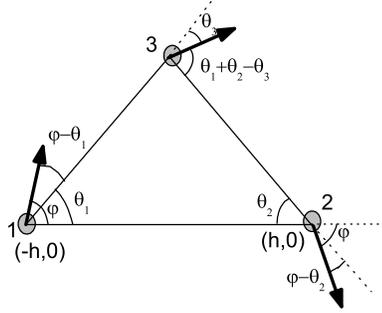}
\end{center}
\caption{Configuration for three dipoles marked by 1, 2, and 3.
The arrows indicate the moments of the dipoles.} \label{fig2}
\end{figure}

\begin{figure}[tbp]
\begin{center}
\includegraphics{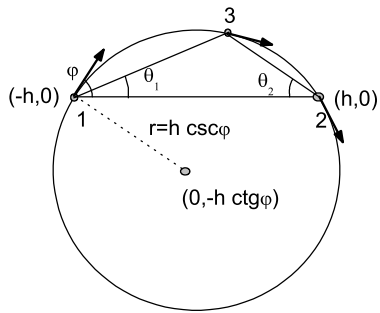}
\end{center}
\caption{For two dipoles 1 and 2 with their centers fixed at $(\pm
h,0)$ and the angles between their moments and the line connecting
their centers to be $\varphi $, the track of the center of dipole
3 to optimize the direction of
its moment is a circle which passes the former two dipoles, centering at $%
(0,-h\cdot ctg\varphi )$, with the radius $h\cdot csc\varphi $. The
directions of the three dipoles are along the tangent of the circle}
\label{fig3}
\end{figure}

\begin{figure}[tbp]
\begin{center}
\includegraphics{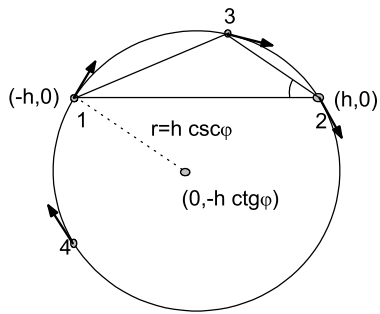}
\end{center}
\caption{When the dipole 4 is added, it should make an optimum
configuration with any two dipoles. As a result, it locates on the
same circle mentioned in Fig.3.} \label{fig4}
\end{figure}

\subsection{Three and four hole-dipoles}

Now let us consider three dipoles case. In Fig.2, three hole-dipoles are
marked by 1, 2, and 3, respectively. Suppose the first and second dipoles
are located at $(\pm h,0)$, with the angles between their dipole moments and
the line connecting them to be $\pm \varphi $. Then we want to find out what
is the optimal location for the third dipole. From Fig. 2, it is easy to see
that for the dipoles marked by 1 and 3, one has
\begin{equation}
\varphi -\theta _1=\theta _3,
\end{equation}
and for 2 and 3, one has
\begin{equation}
\varphi -\theta _2=\theta _1+\theta _2-\theta _3.
\end{equation}
Then one finds
\begin{equation}
\varphi =\theta _2+\theta _1.
\end{equation}
Defining the coordinate of the third dipole by $(x,y)$, we get
\begin{equation}
\varphi =\arctan (\frac y{x+h})+\arctan (\frac y{-x+h}),
\end{equation}
and thus
\begin{eqnarray}
\tan \varphi &=&\tan \left[ \arctan (\frac y{x+h})+\arctan (\frac y{-x+h}%
)\right]  \nonumber \\
&=&\frac{-2hy}{x^2+y^2-h^2},
\end{eqnarray}
such that
\begin{equation}
x^2+(y+\frac h{\tan \varphi })^2=\frac{h^2}{\sin ^2\varphi }.  \label{22}
\end{equation}
>From (\ref{22}), we can see that the track of $(x,y)$ is just a circle
passing through dipole 1 and 2, centered at $(0,-h/\tan \varphi )$ with a
radius $R=h/\sin \varphi $. From the knowledge of geometry, it is easy to
see that the moments of these three dipoles will all point along the
tangents of the circle, as shown in Fig. 3.

Then the energy-minimal configuration for three hole-dipoles is obtained as
an equilateral triangle. The minimal interaction energy for three
hole-dipoles is
\begin{eqnarray}
V_2^{d-d} &\sim &-3\cdot 2q^2\frac{r_0^2}{\left| {\bf r}_{kk^{\prime
}}\right| ^2}  \nonumber \\
&=&-2q^2\frac{r_0^2}{R^2},
\end{eqnarray}
with $\left| {\bf r}_{kk^{\prime }}\right| =\sqrt{3}R$ and $R$ is radius of
the circle crossing three dipoles.

When the fourth hole-dipole is added, it should first make up an optimal
configuration with the dipole 1 and 2, and thus is located on a circle as
discussed above. Then by further making up an optimal configuration with the
dipole 1 and 3, it will also be located on an another circle passing through
the dipole 1 and 3. Since the dipole 3 is already on a circle determined by
1 and 2, with its dipole moment along the tangent, the circle determined by
the dipole 1 and 3 is the same as that determined by the dipole 1 and 2, and
so is the one determined by dipole 2 and 3.

The same argument is applicable to $n$ dipoles. So the optimal configuration
for $n$ dipoles under the potential (\ref{vdd}) will be always on a circle,
with each dipole moment\ along the tangent of the circle and the
proportional spacing for each dipole. Such configuration is just a regular
polygon with $n$ edges. To low energy the radius of the circle $R$ will
shrink until the length of every edge is equal to the minimum $2r_0$,
\begin{equation}
R\simeq \frac{2r_0}{2\sin \frac \pi n}.
\end{equation}
The minimal interaction energy for $n$ hole-dipoles ($n$ is a odd number) is
\begin{eqnarray}
V_n^{d-d} &\sim &-2q^2r_0^2\sum_{kk^{\prime }}\frac 1{\left| {\bf r}%
_{kk^{\prime }}\right| ^2}  \nonumber \\
&=&-2q^2r_0^2\cdot n\cdot 2\cdot \sum_{k=1}^{\frac{n-1}2}\frac 1{\left| {\bf %
r}_k\right| ^2},
\end{eqnarray}
with
\begin{equation}
\left| {\bf r}_k\right| =2R\sin (\frac \pi nk)=2r_0\frac{\sin (\frac \pi nk)%
}{\sin \frac \pi n}.  \label{r_{k}}
\end{equation}
If $n$ is even number, the minimal interaction energy is
\begin{equation}
V_n^{d-d}\sim -2q^2r_0^2n[2\cdot \sum_{k=1}^{\frac n2-1}\frac 1{\left| {\bf r%
}_k\right| ^2}+\frac 12\frac 1{(2R)^2}].
\end{equation}

Numerical simulations for up to $n=30$ dipoles also show that for these
interacting dipoles, the optimized configurations with minimized total
energy are always regular polygons as discussed above.

\begin{figure}[tbp]
\begin{center}
\includegraphics{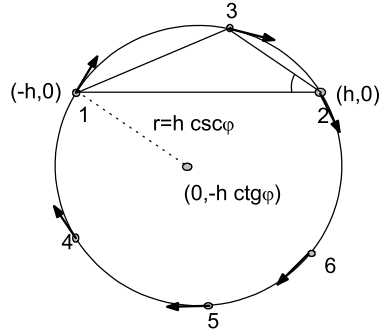}
\end{center}
\caption{When more dipoles are added, they should all locate on
the circle mentioned in Fig.3 and Fig.4. Here $n=6$ is displayed.}
\label{fig5}
\end{figure}

\begin{figure}[tbp]
\begin{center}
\includegraphics{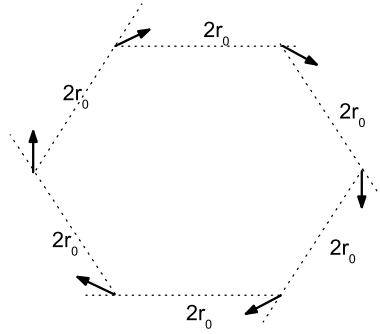}
\end{center}
\caption{The optimum configuration for n dipoles. Their centers
form a regular polygon. The moment of each dipole is directed
along the bisector of each external angle. Here $n=6$ is
displayed.} \label{fig6}
\end{figure}

\begin{figure}[tbp]
\begin{center}
\includegraphics{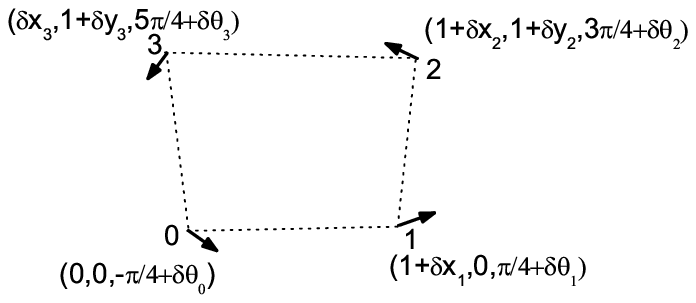}
\end{center}
\caption{A perturbation is imposed to the optimum configuration,
.i.e, the regular polygon. Here $n=4$ is displayed. Each dipole
has a perturbation in the position and direction $(\delta
x_i,\delta y_i,\delta \theta _i)$. The dipole 0 and dipole 1 are
fixed at the origin point and on the x axis respectively, as the
total potential has global $SO(2)$ invariance.} \label{fig7}
\end{figure}

\subsection{Stability of the regular polygon configuration}

In this part we will show the stability of the regular polygon configuration
for hole-dipoles. For a regular polygon configuration, the distance between
two nearest dipoles has reached the minimal $2r_0$. But the distances
between other pairs of dipoles have not arrived at their minimum.
Mathematically, it should be proved that the configuration is stable against
perturbations.

\subsubsection{Perturbation for changing the direction of a dipole moment}

First we change the direction of a dipole moment from it's original
direction $\varphi _0$ to $\varphi _0+\delta \varphi .$ The change of the
total energy for $n$ hole-dipoles ($n$ is a odd number) is
\begin{eqnarray}
\Delta V^{d-d} &\simeq &-2q^2\sum_k\frac{r_0^2}{\left| {\bf r}_k\right| ^2}%
[\cos \left( \varphi _0+\delta \varphi +\varphi _{k^{\prime }}\right) -\cos
\left( \varphi _0+\varphi _{k^{\prime }}\right) ]  \nonumber \\
&\simeq &(q^2\sum_k\frac{r_0^2}{\left| {\bf r}_k\right| ^2})(\delta \varphi
)^2,
\end{eqnarray}
with $\left| {\bf r}_k\right|$ defined by (\ref{r_{k}}). On the other hand,
if $n$ is even number, the energy difference is
\begin{eqnarray}
\Delta V^{d-d}\sim q^2r_0^2[2\cdot \sum_{k=1}^{\frac n2-1}\frac 1{\left|
{\bf r}_k\right| ^2}+\frac 12\frac 1{(2R)^2}](\delta \varphi )^2>0.
\end{eqnarray}

When more hole-dipoles change the directions of their moments
simultaneously, the energy cost is simply the sum of all the positive energy
cost by every change.

So the regular polygon configuration is stable against changing the
directions of the dipole moments $\Delta V^{d-d}>0$.

\subsubsection{Perturbation for changing the positions of the centers of the
dipoles}

Next we change the positions of the centers of the dipoles. In Fig.7, the
case of $n=4$ is shown as an example. The position and direction of each
dipole are marked in Fig.7. Here $2r_0$ is set to be 1 for convenience. The
dipole 0 and dipole 1 are fixed at the origin point and on the x axis
respectively. Each dipole has a perturbation in the position and direction $%
(\delta x_i,\delta y_i,\delta \theta _i)$. After this perturbation, we can
expand the potential up to first order. Taking dipole 1 and dipole 2 as an
example, we have

\begin{equation}
V_{12}=\frac{-V_{0}cos(\varphi _1+\varphi _2)}{r_{12}^2},
\end{equation}
among which

\begin{equation}
V_{0}=2q^2r_0^2.
\end{equation}
For the numerator,we have
\begin{equation}
\varphi _1+\varphi _2=0+\delta \theta _1+\delta \theta _2-2\delta \theta
_{12}.
\end{equation}

In the above equation, $\delta \theta_{1}$ and $\delta \theta_{2}$ denotes
the changes in the directions of the moments of dipole 1 and dipole 2, while
$\delta \theta_{12}$ denotes the change in the direction of line connecting
dipole 1 and dipole 2,
\begin{eqnarray}
\delta \theta_{12}=\delta(x_{1}-x_{2})/1=\delta x_{1}-\delta x_{2}.
\end{eqnarray}
Thus the change in the numerator is zero to the first order of the
perturbation. For the denominator, we have
\begin{eqnarray}
r^{2}_{12}&=&(\delta x_{1}-\delta x_{2})^{2}+(1+\delta y_{2})^{2}  \nonumber
\\
&=& 1+2\delta y_{2}+o(\delta x_{1}^{2},\delta x_{2}^{2},\delta x_{1}\delta
x_{2}).
\end{eqnarray}

As a result, up to the first order perturbation, the potential becomes:
\begin{eqnarray}
V_{12}=\frac{-V_{0}}{1+2\delta y_{2}}=-V_{0}+2V_{0}\delta y_{2}+o(\delta
y^{2}_{2}).
\end{eqnarray}

In the same way, we expanded the potential among other dipoles, and obtained
the following expansion of the total potential up to the first order
perturbation ,
\begin{eqnarray}
\delta V_{tol}^{d-d} &=&\delta (V_{01}+V_{12}+V_{23}+V_{02}+V_{03}+V_{13})
\nonumber \\
&=&\frac {5}{2}V_{0}(\delta x_1+\delta x_2-\delta x_3+\delta y_2+\delta y_3).
\label{dv}
\end{eqnarray}

To satisfy the constraint that the distances between the centers of any two
dipoles should be no less than $2r_{0}$, we expanded the formula of their
distances to the first order perturbation and have:
\begin{eqnarray}
& & \delta x_{1}\geq 0\rightarrow r_{01}\geq 1,  \nonumber \\
& & \delta x_{2}\geq \delta x_{3}\rightarrow r_{23}\geq 1,  \nonumber \\
& & \delta y_{2}\geq 0\rightarrow r_{12}\geq 1,  \nonumber \\
& & \delta y_{3}\geq 0\rightarrow r_{03}\geq 1.  \label{constraint}
\end{eqnarray}

>From (\ref{dv}) and (\ref{constraint}), we obtained that:
\begin{equation}
\delta V_{tol}\geq 0,  \label{dv4}
\end{equation}
which denotes that the change of the total potential is no less than zero up
to the first order perturbation.

Our above demonstration can be easily generalized to arbitrary $n$. For a
general $n$, we have
\begin{equation}
V_{tot}=\sum_{ij}\frac{-V_{0}cos(\varphi _i+\varphi _j)}{r_{ij}^2}.
\end{equation}
Up to the first order perturbation, we have:
\begin{equation}
\delta V_{tot}=V_{0}\sum_{ij}\frac{2(x_{i}-x_{j})(\delta x_{i}-\delta
x_{j})+2(y_{i}-y_{j})(\delta y_{i}-\delta y_{j})}{r_{ij}^4}.
\end{equation}
The constraint for the least distance between any two adjacent dipoles
reads:
\begin{equation}
\delta r_{i,i+1}=2(x_{i}-x_{i+1})(\delta x_{i}-\delta
x_{i+1})+2(y_{i}-y_{i+1})(\delta y_{i}-\delta y_{i+1})\geq 0.  \label{dr}
\end{equation}

It can be checked that
\begin{equation}
\delta V_{tot}=gV_{0}\sum_{i}\delta r_{i,i+1},  \label{dvt}
\end{equation}
in which
\begin{equation}
g=\frac{\sum_{i=2}^{n-1}\sin[\pi(i-1)/n]/[\sin(\pi i)/n]^3}{%
8\sin(\pi/n)\sin(2\pi /n)}>0.  \label{g}
\end{equation}

>From (\ref{dr}), (\ref{dvt}) and (\ref{g}), we have
\begin{equation}
\delta V_{tot}\geq 0 .  \label{dvn}
\end{equation}
>From (\ref{dvn}), we know that the change of the total potential is no less
than zero up to the first order perturbation for arbitrary $n$. When all the
$\delta x_{i}$ and $\delta y_{i}$ are carefully chosen so that the "=" is
realized in (\ref{dr}), (\ref{dvn}) turns into
\begin{equation}
\delta V_{tot}=0 .  \label{dvn1}
\end{equation}
In such cases, we should have to check the second order perturbation in the
potential energy as the first order perturbation is zero. We again gave our
proof for $n=4$ as an example.

When $\delta x_1=\delta y_2=\delta y_3=0$ , and $\delta x_2=\delta
x_3=\delta x$, the "=" is realized in (\ref{constraint}) and hence (\ref{dv4}
). The first order perturbation of the total potential is zero, so we
expanded it to the second order perturbation, and obtained
\begin{eqnarray}
V_{tol} &\geq &\sum_{i,j}\frac{-V_{0}}{r_{ij}^2}  \nonumber \\
&=&-2V_{0}-\frac {2V_{0}}{1+(\delta x)^2}-\frac {V_{0}}{1+(1-\delta x)^2}-%
\frac {V_{0}}{1+(1+\delta x)^2}  \nonumber \\
&=&-5V_{0}+\frac {3}{2}V_{0}(\delta x)^2+o((\delta x)^3),
\end{eqnarray}
and therefore,
\begin{equation}
\delta V_{tol}\geq \frac {3}{2}V_{0}(\delta x)^2\geq 0.
\end{equation}
So the change in the total potential is also no less than zero up to the
second order perturbation.

In the case of $n>4$, we can carry out similar expansion, and draw the same
conclusion.

By the above perturbative expansion, we proved that the configuration shown
in Fig. 6 is stable against local perturbation in the positions and
directions of the dipole moments.

\subsection{Energetically minimized configurations for infinite number
hole-dipoles - stripes}

In the above part, we have proved that $n$ dipoles will arrange themselves
to form a regular polygon to minimize the energy. And the radius of the
circle $R$ will shrink until the length of every edge is equal to the
minimum $2r_0$. When $n\rightarrow \infty $, the radius of the circle $R$
turn to diverge $R\simeq \frac{r_0}{\sin \frac \pi n}\sim \frac{r_0}\pi
n\rightarrow \infty $ and a regular polygon will naturally be stretched into
a line ({\em i.e.,} a stripe), as shown in Fig. 8. The minimal interaction
energy for a regular polygon with large number of hole-dipoles is about
\begin{eqnarray}
V^{d-d} &\sim &-2q^2r_0^2\sum_{kk^{\prime }}\frac 1{\left| {\bf r}%
_{kk^{\prime }}\right| ^2}  \nonumber \\
&\simeq &-q^2n
\end{eqnarray}
When a dipole leaves away from the line ( a stripe), finite energy $\Delta
V^{d-d}$ will be cost
\begin{eqnarray}
\Delta V^{d-d} &\simeq &-2q^2r_0^2\sum_{k^{\prime }}\frac 1{\left| {\bf r}%
_{k^{\prime }}\right| ^2}+2q^2r_0^2\sum_k\frac 1{\left| {\bf r}_k\right| ^2}
\nonumber \\
&=&\frac{q^2\pi ^2}{24}(\delta r)^2,
\end{eqnarray}
where ${\bf r}_k=2kr_0$ and $\left| {\bf r}_{k^{\prime }}\right| ^2=\left| 2%
{\bf r}_k\right| ^2+(\delta r)^2$. Thus the line shape configuration -
stripe is stable against local perturbation.

It is easy to see that such a ``stripe''\ of charge carriers is embedded in
a domain wall of the AF background. To see this, we consider a stripe along
the $\hat{x}$-axis composed of the hole-dipoles of a size $\left| {\bf e}%
\right| =r_0$ and spaced by $l=\alpha \left| {\bf e}\right| $. Far away from
the $\hat{x}$-axis, the total spin twist summed from (\ref{dipole}) is given
by
\begin{eqnarray}
\phi _i &=&\sum_k\phi _i^k=\sum_k\frac{e_xy_{ik}}{r_{ik}^2}  \nonumber \\
&\simeq &\frac \pi \alpha \text{sgn}(y_i),
\end{eqnarray}
when $\left| y_i\right| \gg r_0\sim a.$ Here $y_{ik}=y_i$ and $x_{ik}=lk+x_i$
according to the definition. Thus a phase shift is found across the stripe
with
\begin{equation}
\Delta \phi =\phi _{y>0}-\phi _{y<0}=2\cdot \frac \pi \alpha .
\end{equation}

For a special case, $\alpha =2$, the line becomes an antiphase domain wall
\begin{equation}
\Delta \phi =\phi _{y>0}-\phi _{y<0}=2\cdot \frac \pi \alpha =\pi .
\label{deltaphi}
\end{equation}
Namely, a stripe composed of hole-dipoles is topologically an antiphase
domain wall.

>From (\ref{deltaphi}), we can also understand physically the reason why the
holes tend to arrange themselves into a straight line. When the hole-dipoles
are distributed randomly in the AF background, each dipole induces a spin
twist as described by (\ref{dipole}), which costs additional energy. But
when the hole-dipoles form a stripe, the total twist spins away from the
domain wall will be cancelled out such that the spin ordering on either side
of the domain wall becomes unfrustrated, just like at half-filling.

Yet there is a further advantage in the formation of the stripe, {\em i.e,}
the kinetic energy of the holes can be gained. Recall that an isolated
hole-dipole cannot move freely as it is self-trapped in real space. But when
the hole-dipoles arrange themselves into a straight line, the individual
holes actually may move freely along the stripe such that a delocalization
energy can be gained. This will correspond to a metallic stripe case.

With a further increase of doping, more stripes will be formed as
shown in Fig. 9. For any two adjacent stripes, without the spin
frustration inside the domain between them, they will not gain
additional energy by being closer. Under a long-range Coulomb
repulsion, a uniform stripe phase will be stable against a cluster
formation. In this uniform stripe phase, the distances between two
neighboring stripes are the same, which obviously is determined by
the hole concentration. It should be noticed that the recently
found "checkerboard" pattern in the LDOS by the STM in the cuprate
may display a new kind of CDW order\cite{checkerboard}. Its
possibility in the framework of the above mentioned dipole picture
is under further exploration when the long-range Coulomb repulsion
is considered more carefully.

\begin{figure}[tbp]
\begin{center}
\includegraphics{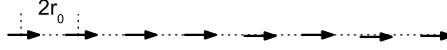}
\end{center}
\caption{When $n\rightarrow \infty $, the regular polygon shown in
Fig.6 turns into a line. The distance between the centers of
neighboring dipoles is $2r_{0}$.} \label{fig8}
\end{figure}

\begin{figure}[tbp]
\begin{center}
\includegraphics{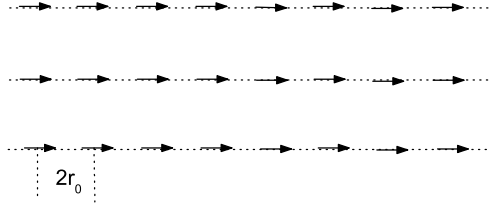}
\end{center}
\caption{When more holes are doped to occupy a certain
concentration of the lattice grid, parallel lines are formed to
make the stripe phase} \label{fig9}
\end{figure}

\section{The effects of Zn-impurities}

As emphasized before, the stripe formation is intimately related
to the self-localization of individual hole-dipoles. We also have
mentioned that disorder or impurities in the system may have an
\textquotedblleft amplified\textquotedblright\ effect on
localization.

In the following we discuss the effect of Zn impurities on hole-dipoles as
well as the stripe phase. It is well known that doped Zn atoms will be
present in the form of ${\rm Zn}^{2+}$ with a closed-shell structure,
substituting the ${\rm Cu}^{2+}$ sites in the ${\rm CuO}_{2}$ planes of the
cuprates. In the t-J model, the site occupied by the ${\rm Zn}^{2+} $ may be
imposed by a boundary condition of an \textquotedblleft
empty\textquotedblright\ site where no electron or hole can stay there at
low energy.

Let us examine how a ${\rm Zn}$ impurity and a hole-dipole will interact.
Inside a hole-dipole, those spins on a loop circling the center of the
antimeron will have a $2\pi $ rotation in their polarization directions. As
the radius of such a loop shrinks continuously to the antimeron core, the
spin polarization directions will change quickly and becomes uncertain at
the core site. So a spin at the core site of the antimeron will just like a
"defect" spin \cite{kou1}, and the bonds which connect the core spin with
its surrounding spins can be thus viewed to be\ effectively
\textquotedblleft cut off\textquotedblright , resulting an energy increase
roughly $\sim 4J^{\prime }$ ($J^{\prime }$ denotes the average superexchange
energy for one bond). On the other hand, when a ${\rm Zn}^{2+}$ is doped and
replaces a normal ${\rm Cu}^{2+}$ site, the bonds which used to connect such
a ${\rm Cu}^{2+}$ with its surroundings are cut off with an energy cost
approximately $\sim 4J^{\prime }$. But, if such a ${\rm Zn}^{2+}$ is to
replace the ${\rm Cu}^{2+}\,$ sitting at the core of the antimeron of a
hole-dipole, no additional superexchange energy will be cost by breaking up
those four bonds connected to the impurity site. Therefore, it will be
energetically favorable for a hole-dipole to be trapped by a ${\rm Zn}$
impurity. As shown in Fig.10, a Zn impurity locates at the center of the
antimeron. Experimentally, there is evidence \cite{exp} that doped holes are
indeed trapped by the Zn impurities.

\begin{figure}[tbp]
\begin{center}
\includegraphics{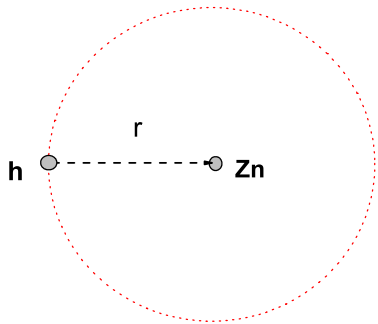}
\end{center}
\caption{When a hole and a $Zn^{2+}$ are doped into the AF
background
simultaneously, they would like to attract each other and form a $Zn^{2+}$%
-holon dipole. H denotes the holon, while Zn denotes the $Zn^{2+}$}
\label{fig10}
\end{figure}

\begin{figure}[tbp]
\begin{center}
\includegraphics{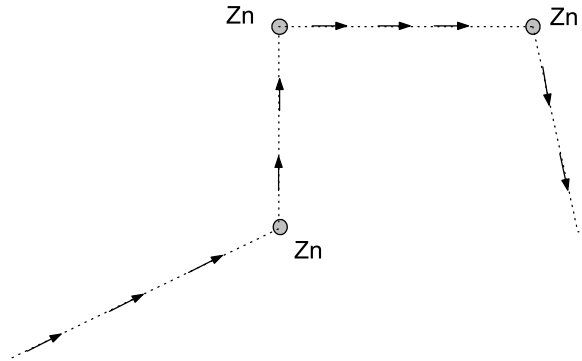}
\end{center}
\caption{A stripe in the presence of the $Zn^{2+}$. The stripe has
to pass the $Zn^{2+}$, and so is pinned.} \label{fig11}
\end{figure}

According to the above discussions, a large number of Zn impurities will not
favor the stripe formation, for they tend to trap hole-dipoles around
themselves. The random distribution of the Zn leads to the random
distribution of the hole-dipoles. If the Zn concentration is very low, then
a stripe is expected to be easily pinned by a Zn, and be bent in order to
pass several zincs, as shown in Fig.11. From the transport measurements\cite
{transport} and also from the muon-spin- relaxation ($\vert$\`{I}SR)
measurements\cite{usr2}, it is found that a small amount of Zn impurities
are effective for the pinning. So we predict that Zn impurities are very
effective in destroying a stripe phase at low doping.

\section{Conclusion}

In this paper, it is shown that in the framework of the phase-string model,
each hole doped into a spin ordered phase at low doping will act as a dipole
which is self-trapped in real space. With the suppression of the kinetic
energy, the dipole-dipole interaction between hole-dipoles will dominate the
low-energy physics in the absence of disorder or impurities. We demonstrated
in detail how a few hole-dipoles collapse into a stable configuration of the
regular polygon, which turns into a stripe (stripes) in thermodynamic limit.
Consequently, we found that at a finite concentration of holes at low
doping, a uniform stripe phase is highly competitive. The effects of
impurities are also discussed. When a ${\rm Zn}^{2+}$ is doped into the
system , it generally tends to trap a hole around itself to form a ${\rm Zn}%
^{2+}$-holon dipole. As a result, the stripe will be pinned near the ${\rm Zn%
}^{2+}$ site. Further more, we predicted that a finite concentration of
zincs can easily kill a uniform stripe phase.

\begin{acknowledgments}
We acknowledge stimulating discussions with Z. Y. Weng. This work is partially supported by NSFC grant no. 90103021, no. 10247002, and no. 10204004. S. P. Kou acknowledges the support from Beijing Normal University.

\end{acknowledgments}

\end{document}